\documentclass[twocolumn,
               superscriptaddress]{revtex4}
\usepackage{graphicx,latexsym}

\begin{document}

\title{\Large\bf $\bf\Lambda(1405)$ and Negative-Parity Baryons in Lattice QCD}

\author{Y.~Nemoto}
\affiliation{RIKEN-BNL Research Center, Brookhaven National Laboratory,
 Upton, NY 11973, USA}

\author{N.~Nakajima}
\affiliation{Center of Medical Information Science,
             Kochi University, Kochi, 783-8505 Japan}

\author{H.~Matsufuru}
\affiliation{Computing Research Center, High Energy Accelerator Research
Organization (KEK), Tsukuba 305-0801, Japan}

\author{H.~Suganuma}
\affiliation{Faculty of Science, Tokyo Institute of Technology,
             Tokyo 152-8551, Japan}


\begin{abstract}
We review briefly 
recent studies of the $\Lambda(1405)$ spectrum in Lattice QCD.
Ordinary three-quark pictures of the $\Lambda(1405)$ in quenched Lattice
QCD fail to reproduce the mass of the experimental value, 
which seems to support the penta-quark picture for the $\Lambda(1405)$ 
such as a $\bar{K}N$ molecule-like state.
It is also noted that the present results suffer from relatively large
systematic uncertainties coming from the finite volume effect, the 
chiral extrapolation and the quenching effect.
\end{abstract}

\maketitle

\section{Introduction}

 The $\Lambda(1405)$ is still a mysterious particle due to its small
mass.
It is the lightest negative-parity baryon although it contains a strange
valence quark.
The conventional quark model, in which the $\Lambda(1405)$ is assigned
as the flavor-singlet state in the 70 dimensional representation under
the spin-flavor SU(6) symmetry, cannot explain this small mass.
Since the mass is just below the $\bar{K}N$ threshold,
there is another interpretation based on a
five-quark picture, a $\bar{K}N$ bound state or a $\pi\Sigma$ resonance.
It is likely that the actual $\Lambda(1405)$ will be a mixed state of these
two pictures.
However, knowing which picture is the dominant contribution is quite important
for hyper-nuclear physics and astro-nuclear physics, because the proton in 
the $K^- p$ bound state suffers from the Pauli blocking effect in nuclear
matter.
It may influence on the existence of the kaonic nuclei and kaon
condensation in neutron stars.

Quenched lattice QCD is a useful tool to distinguish between the three- and
the five-quark pictures of the $\Lambda(1405)$, because it is more 
valence-like than dynamical QCD due to the absence of sea quarks.
Recently we have calculated the $\Lambda(1405)$ and other low-lying
negative-parity baryon spectra in anisotropic quenched lattice QCD
\cite{nem03}.
In this talk we mainly report our results and
comment on other recent two lattice studies on the $\Lambda(1405)$ 
spectrum \cite{mel03, lee02}.

\section{Lattice formulation}

We employ the standard Wilson gauge action and the $O(a)$ improved Wilson
quark action with tadpole improvement.
We adopt the anisotropic lattice since fine resolution in the temporal
direction makes us easy to follow the change of heavy-baryon correlations 
and to determine the fitting ranges for their masses.
We take the renormalized anisotropy as $\xi=a_\sigma/a_\tau=4$,
where $a_\sigma$ and $a_\tau$ are the spatial and temporal lattice
spacings, respectively.
The simulation is carried out on the three lattices 
where parameters are well tuned and errors are rather well evaluated \cite{mat01}.
The sizes of the lattices are 
$12^3\times96 (\beta=5.75,$ i.e., $a_\sigma^{-1}=1.034(6)$GeV),
$16^3\times128 (\beta=5.95,$ i.e., $a_\sigma^{-1}=1.499(9)$GeV),
and
$20^3\times160 (\beta=6.10,$ i.e., $a_\sigma^{-1}=1.871(14)$GeV).
The scale $a_\sigma^{-1}$ is determined from the $K^*$ meson mass.
Determination of the other parameters is described in Ref.\cite{mat01}.

As for the quark mass, we adopt four different values 
which roughly cover around strange quark mass and 
correspond to the pion mass being about $0.6-0.9$GeV. 
(See also Fig. \ref{res}). 
We use the standard baryon operators 
which survive in the nonrelativistic limit: 
the $(\bar{q}^TC\gamma_5)q$ form for the octet 
and the $(\bar{q}^T C\gamma_\mu)q$ form for the decuplet.
Here $C$ is the charge conjugation matrix.
For the $\Lambda(1405)$ we use the flavor-singlet operator following
the assignment of the quark model,
\begin{equation}
  (u^TC\gamma_5d)s+(d^TC\gamma_5u)s+(s^TC\gamma_5u)d,
\end{equation}
where we have omitted the color indices for simplicity.
Both the positive- and negative-parity baryon spectra can be obtained
from these operators using the parity projection, because they
couple to both the parity states.
In the source operator, each quark is spatially smeared with the
Gaussian function with the width $\sim0.4$ fm for better overlap
with the low-lying states.

\section{Numerical results}

\begin{figure*}
\includegraphics[width=8.8cm]{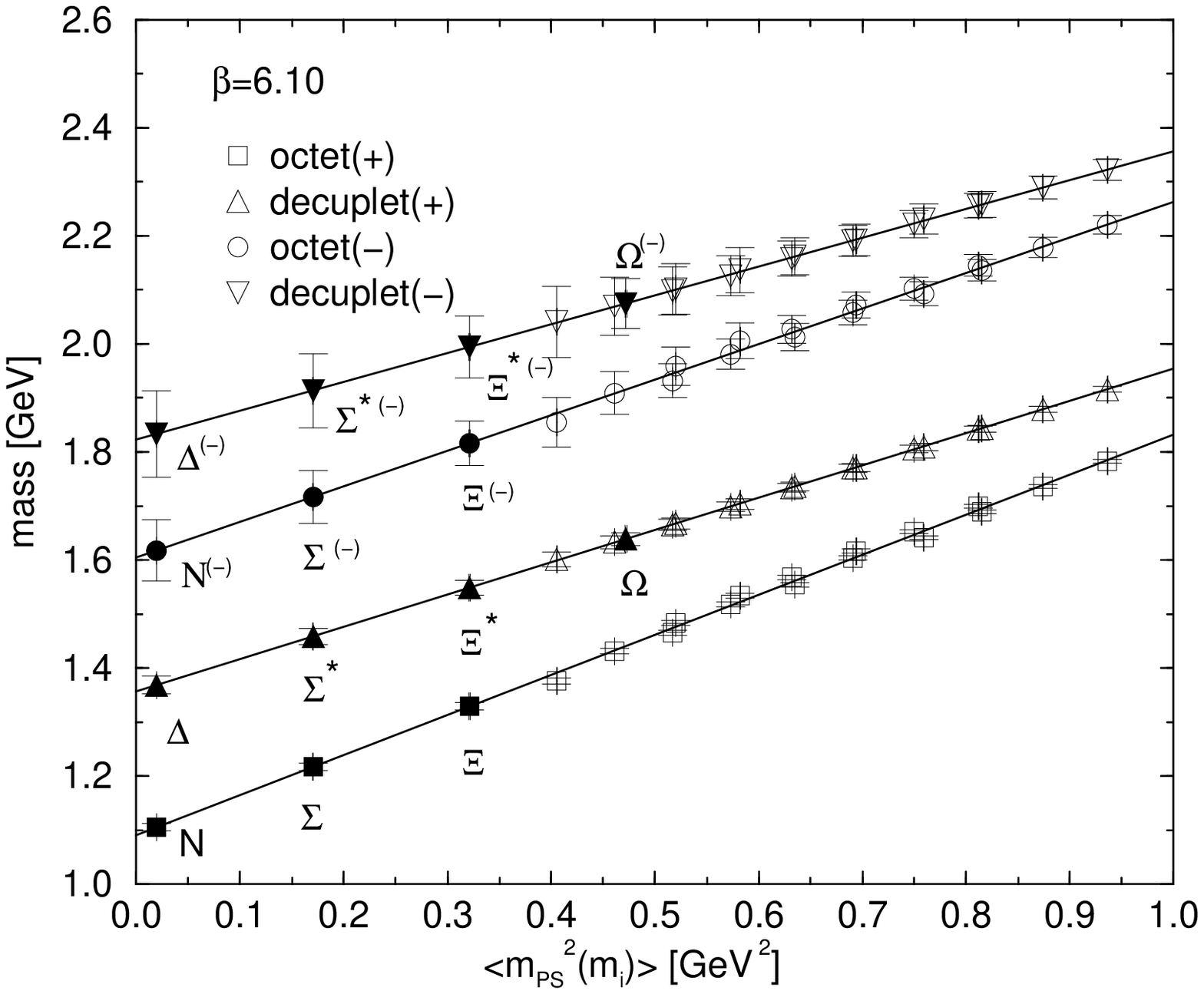}
\includegraphics[width=8.8cm]{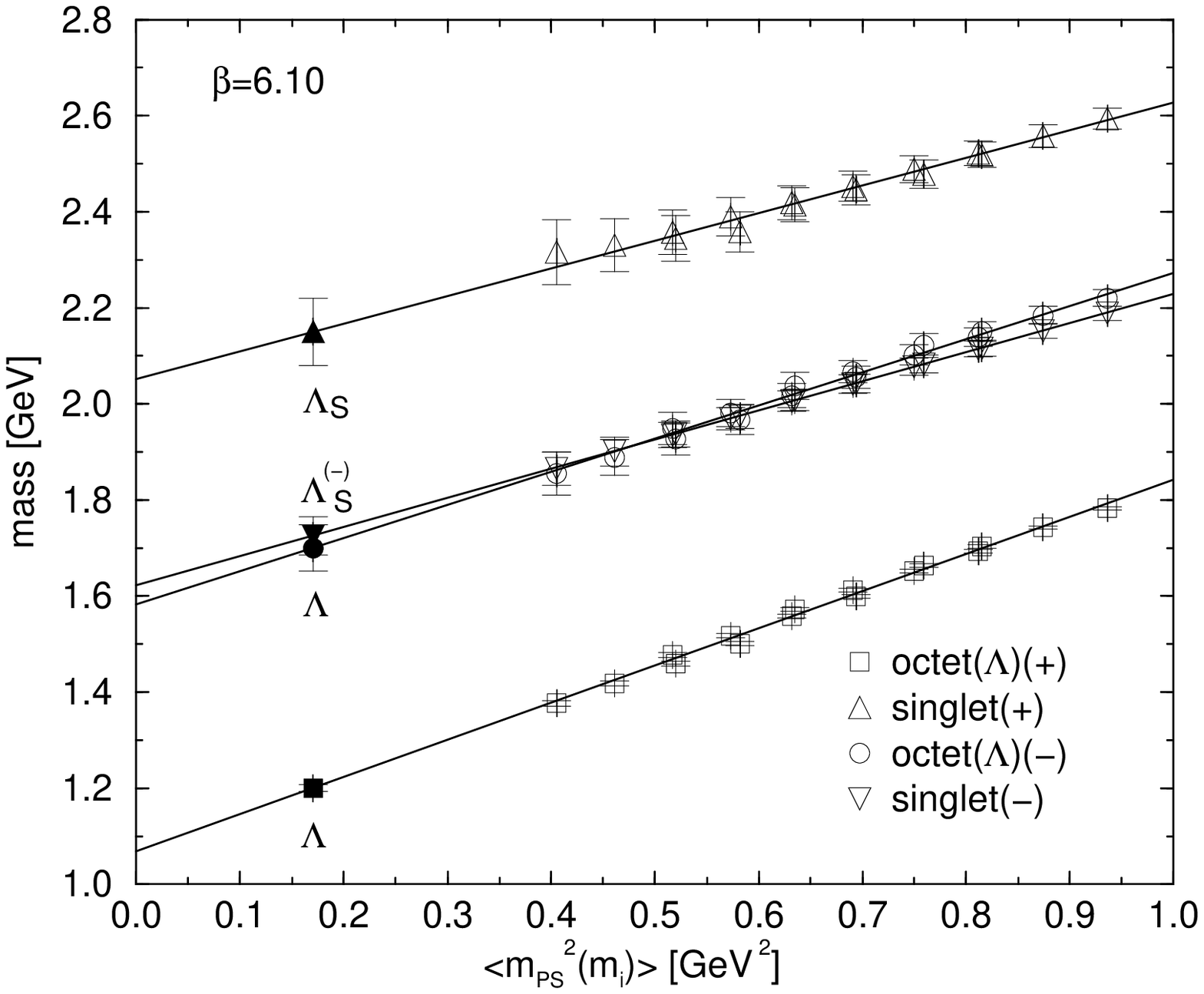}
\caption{The positive- and negative-parity baryon spectra for the
$\beta=6.10 (a_\sigma^{-1}\simeq 1.9$GeV) lattice.
The octet and the decuplet baryons are shown in the left figure and the 
octet($\Lambda)$ and the singlet baryons in the right.
The open symbols denote the lattice data and the filled symbols
the fitted results from the linear chiral extrapolation.}
\label{res}
\end{figure*}

The numerical results and the fit results of the baryon spectrum
for the finest $\beta=6.10 (a_\sigma^{-1}\simeq 1.9$GeV) lattice are shown in
Fig.\ref{res}. 
The results from the other lattices are similar \cite{nem03}.
The physical $u$, $d$ and $s$ quark masses are determined with the
$\pi$ and $K$ meson masses.
For each baryon, two of quark masses are taken to be the same value,
$m_1$, and the other quark mass $m_2$ is taken to be an
independent value.
The baryon masses are then expressed by the function of $m_1$ and
$m_2$, but the numerical results seem to be well described with
the linear form, $m_B(m_1,m_2)=m_B(0,0)+B_B\cdot (2m_1+m_2)$.
Therefore we fit the baryon spectrum to the linear form in the
sum of corresponding pseudoscalar meson mass squared,
$\langle m_{\rm PS}^2(m_i)\rangle 
= 1/N_q \sum_{i=1}^{N_q}m_{\rm PS}^2(m_i,m_i) = 2B \sum_{i=1}^{N_q} m_i
$ with $N_q=3$ for
baryons, as shown in Fig. \ref{res}.

We have also calculated the vector meson spectrum with the same way 
mentioned above and have used the same fitting method.
Our results for the vector meson and the positive-parity baryons
are consistent with those obtained in Ref. \cite{mat01}.

\section{Discussions}

From Fig.\ref{res}, the chiral extrapolated results of the 
negative-parity baryon spectrum are heavier than those of the
corresponding positive-parity sectors for octet and decuplet,
 as expected.
The flavor-singlet negative-parity baryon is, however, lighter
than the positive-parity one.
This is consistent with the conventional quark model,
in which the flavor-singlet positive-parity baryon is assigned
as a state with the principal quantum number $N=2$, while 
the negative-parity sector has $N=1$.

\begin{figure*}
\includegraphics[width=8.3cm]{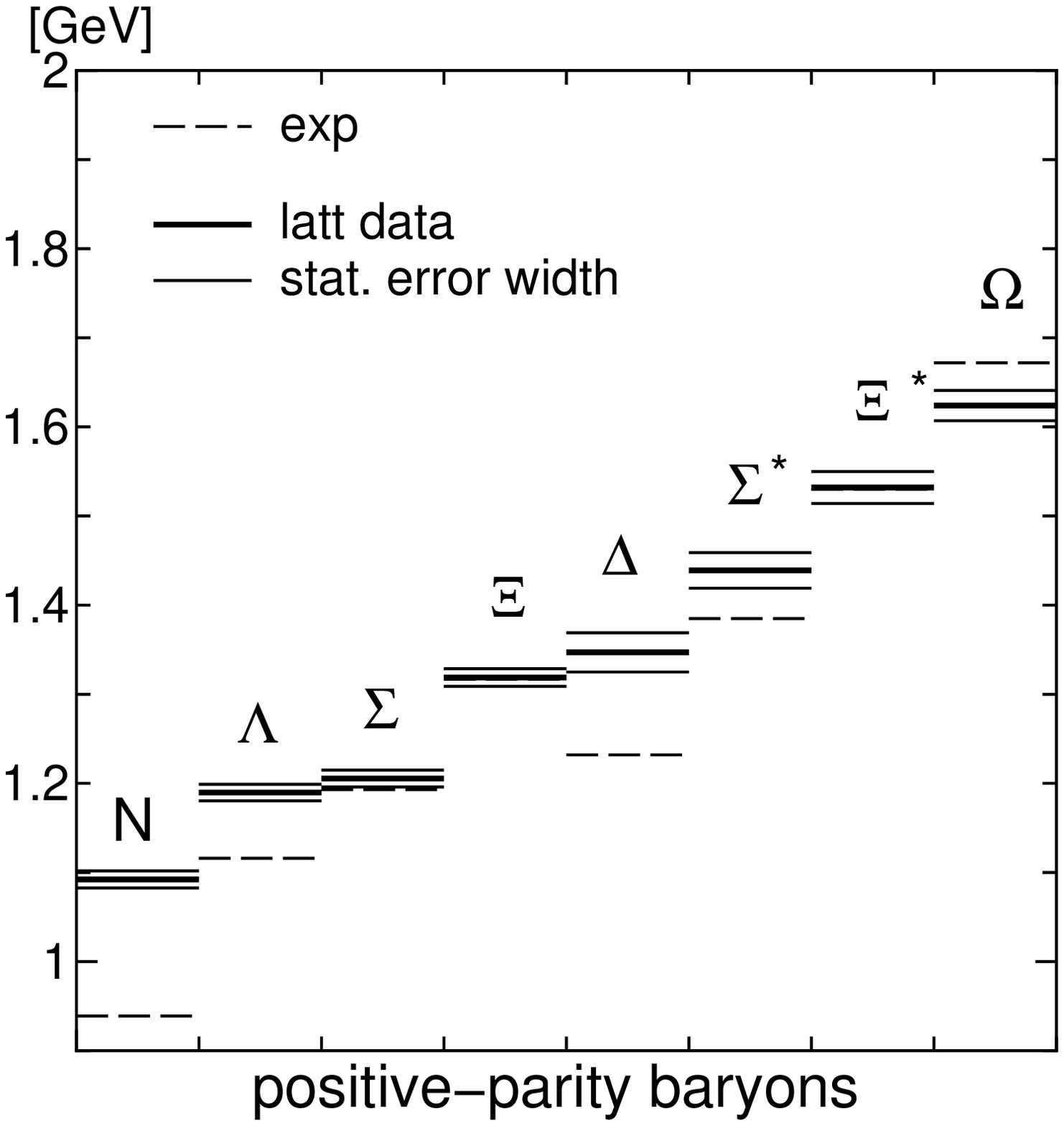}
\includegraphics[width=8.8cm]{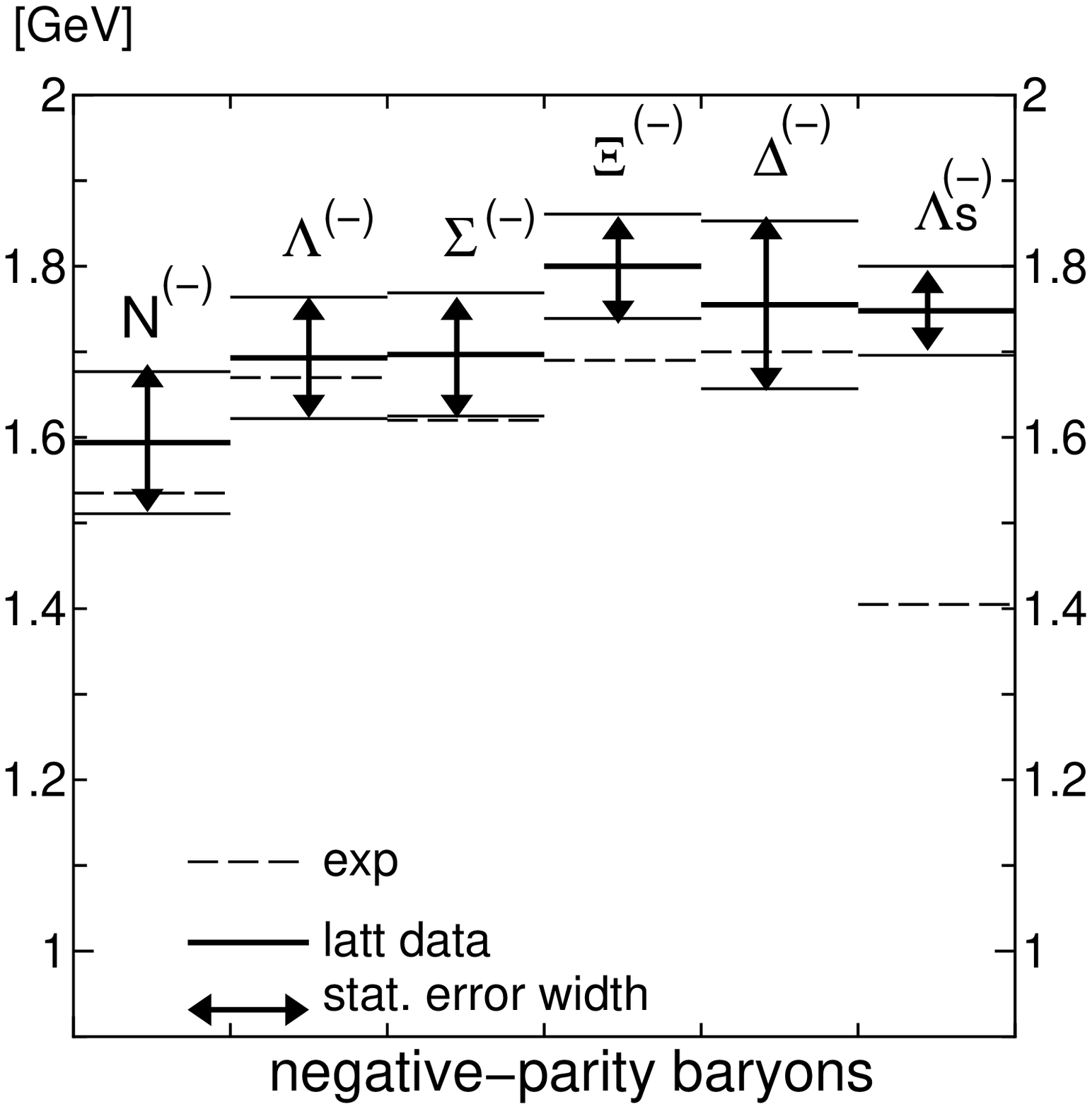}
\caption{Calculated and experimental baryon masses. The lattice results
are taken from the $\beta=6.10$ lattice. For the negative-parity baryons,
the experimental values are $N(1535), \Lambda(1670), \Sigma(1620),
\Xi(1690), \Delta(1700)$ and $\Lambda(1405)$.}
\label{comp}
\end{figure*}

Various baryon masses at the $\beta=6.10$ lattice together with
the experimental values are shown in Fig.\ref{comp}.
As for the negative-parity baryons, most of the lattice 
results comparatively well reproduce the experiment.
The flavor-singlet baryon is, however, exceptional:
its calculated mass of about 1.7GeV is much heavier than
the $\Lambda(1405)$ with a difference of more than 300 MeV.
The difference between them is actually the largest in all
the hadrons in consideration.
We also note that the flavor-singlet baryon has one strange
valence quark and therefore the systematic error from the
chiral extrapolation should be less than that of the nucleon
and the delta.
Even if one takes the systematic error coming from the quenching effect 
of 10\% level into account, this discrepancy of more than 300 MeV
cannot be accepted.
Therefore it is natural to consider the flavor-singlet baryon is
physically different from the $\Lambda(1405)$.
The flavor-octet negative-parity $\Lambda$ baryon mass is almost
the same as the flavor-singlet one and thus it also is not likely to be
a candidate of the $\Lambda(1405)$.

We turn to other two lattice studies on the $\Lambda(1405)$.
Melnitchouk {\it et al.} investigated excited state baryons using
an improved quenched Wilson gauge and an improved Wilson quark actions
\cite{mel03}.
They computed an operator composed of the terms common to the flavor-octet 
$\Lambda$ operator
and the flavor-singlet operator, i.e.,
\begin{equation}
  (d^T C\gamma_5 s)u + (s^T C\gamma_5 u)d
  \label{common}
\end{equation}
to study the $\Lambda(1405)$.
This allows for mixing between octet and singlet states by the favor
SU(3) symmetry breaking due to the strange quark mass.
Although they did not take the chiral extrapolation of the results,
the result for the common operator is still much heavier than the
$\Lambda(1405)$, about 1.8GeV, 
if one assumes the simple linear form for the chiral extrapolation.
Therefore, as their conclusion, the $\Lambda(1405)$ does not couple strongly
 to the operator (\ref{common}) and further investigations such as
with lighter quark masses or dynamical quarks are needed.

Lee {\it et al.} studied excited state baryons using the overlap
fermions \cite{lee02}.
They did not employ any special operators for the $\Lambda(1405)$ but
the usual flavor-octet operators to compare with experiment.
The overlap fermion is able to access light quark masses than ever
because of the good chiral property.
In fact they took very light quark masses for which the lightest pion 
mass becomes about 0.18GeV.
The quark mass dependence of the baryons is apparently different from
the other two studies:
non-linear forms of the chiral extrapolation are clearly seen in the
light quark mass region $(m_\pi^2<0.3 {\rm GeV}^2$).
The flavor-octet negative-parity spectrum rapidly decreases leading
to about 0.14GeV as the quark mass decreases.
(Our result \cite{nem03} and the result of Ref.\cite{mel03} do not show 
such behavior, because their pion masses are about 0.6GeV at most.)
Such a light quark mass region can, however, suffer from large systematic
uncertainties even in the chiral sophisticated fermions.
In particular some ghost effects coming from unitarity violation
of the quenched approximation may be significant in this region.
Also a rather coarse lattice ($a^{-1}\sim 1{\rm GeV}$) which they used would  
give hadron spectrum an additional uncertainty.
Hence further systematic error analyses are needed also in this case.

\section{Summary}

We have focused on the $\Lambda(1405)$ spectrum in quenched lattice QCD.
In our calculation with the anisotropic lattice, 
the flavor-singlet baryon mass measured with the three-quark operator is found to be about 1.7GeV, 
which is much heavier than the $\Lambda(1405)$.
In Ref.\cite{mel03}, an operator with terms common to the flavor-octet
$\Lambda$ and the flavor-singlet states has been used to describe the
$\Lambda(1405)$, and 
the result seems to be very heavy after the naive chiral 
extrapolation.
Both the results present the difficulty in identifying the $\Lambda(1405)$ as the flavor-singlet three-quark baryon, 
which seems to indicate that the $\Lambda(1405)$ is dominated by 
an exotic state such as a $\bar{K}N$ molecule-like state.

On the other hand, it is reported in Ref.\cite{lee02} that
the linear chiral extrapolation of the excited baryon spectrum
fails in the very light quark mass region using overlap fermions.
Their results are closer to experiment than the others, while
further systematic error analyses coming from the quenching effects
are needed in that light quark region.

Thus for more definite understanding of the $\Lambda(1405)$ on the lattice,
it is desired to carry out more extensive work such as a simulation with
dynamical quarks or pentaquark states like the $\bar{K}N$ state.

Y.N. thanks the hospitality at the Yukawa Institute for Theoretical Physics at 
Kyoto University and fruitful discussions during the YITP Workshop YITP-W-03-21 
on ``Multi-quark Hadron: four, five and more?".


\thebibliography{9}
\bibitem{nem03}
  Y. Nemoto, N. Nakajima, M. Matsufuru, H. Suganuma, Phys. Rev. D {\bf 68},
 094505 (2003); Nucl.\ Phys.\ {\bf A721}, 879 (2003); 
  N. Nakajima, M. Matsufuru, Y.Nemoto and H.~Suganuma, 
  AIP Conf. Proc. {\bf CP594}, 349 (2001). 
\bibitem{mel03}
  W.~Melnitchouk {\it et al.}, Phys.\ Rev.\ D {\bf 67}, 114506 (2003).
\bibitem{lee02}
  F.~X.~Lee, S.~J.~Dong, T.~Draper, I.~Horvath, K.~F.~Liu, N.~Mathur and 
  J.~B.~Zhang, Nucl.\ Phys. {\bf B} (Proc.\ Suppl.) {\bf 119}, 296 (2003).
\bibitem{mat01}
H.~Matsufuru, T.~Onogi and T.~Umeda,
  Phys.\ Rev.\ D {\bf 64}, 114503 (2001).
\end{document}